\documentclass[aps,prl,a4paper,notitlepage,reprint,superscriptaddress]{revtex4-1}

\usepackage{amsmath,amssymb,amsfonts} 
\usepackage{mathrsfs} 
\usepackage{upgreek}
\usepackage{mathtools} 
\usepackage{color}
\usepackage{graphicx,float}
\usepackage[colorlinks,
linkcolor=red,
citecolor=blue,
urlcolor=red]{hyperref}

\usepackage{xcolor}    
\usepackage{soul}
\usepackage{subcaption}

\usepackage{ragged2e} 

\usepackage{changes}
\definechangesauthor{zhao}
\setdeletedmarkup{\textcolor{red}{\sout{#1}}}
\setaddedmarkup{\textcolor{blue}{\uwave{#1}}}

\newcommand{\figref}[1]{Fig.~\ref{#1}}
\renewcommand{\eqref}[1]{Eq.~\ref{#1}}

\begin{document}

\title{Predictive Model for the Dissipation Dilution of Soft-Clamped Modes in Polygon Resonators}

\author{Zhihao Niu}
\email{niuzhihao@quantumsc.cn}
\affiliation{Quantum Science Center of Guangdong-Hong Kong-Macao Greater Bay Area, Shenzhen 518045, China}
\affiliation{Modern Matter Laboratory and Advanced Materials Thrust, The Hong Kong University of Science and Technology (Guangzhou), Guangzhou 511400, China}
\author{Yuanyuan Zhao}
\email{zhaoyuanyuan@quantumsc.cn}
\affiliation{Quantum Science Center of Guangdong-Hong Kong-Macao Greater Bay Area, Shenzhen 518045, China}
	
\begin{abstract}
Polygon resonators are promising candidates for nanomechanical applications due to their compact architecture and high force sensitivity. Here, we develop a polygon-specific analytical model for perimeter modes by coupling the flexural motion of the polygon sides to tether torsion governed by a stress-modified Timoshenko-Gere (TG) equation. The model captures finite-wavelength torsional phase accumulation beyond the conventional short-tether approximation, revealing tether-length-dependent resonances and bending-torsion hybridization. Comparisons with finite-element-method (FEM) simulations show good agreement in the quasi-one-dimensional regime, while identifying the onset of deviations associated with neglected torsional deformation of the polygon sides. The model therefore provides a useful analytical tool for interpreting and optimizing high quality factor ($Q$)
polygon resonators for cavity optomechanics and precision sensing.

\end{abstract}

\date{\today}
\maketitle
	 

\paragraph{Introduction. ---} 
Recent advancements have facilitated ultra-high quality factor ($Q$) nanomechanical resonators, attributed to the phenomenon known as "dissipation dilution" \cite{tsaturyan2017ultracoherent,gonzalez2000suspensionsthermal}. This phenomenon originates from the interplay between geometric nonlinearity of strain in deformations and static strain \cite{fedorov2019generalized}. Notably, in stress-dominated resonators, dissipation dilution can strongly suppress mechanical energy loss through geometric and strain engineering, reducing the reliance on material optimization alone.

Boundary-localized mode curvature is a major source of mechanical dissipation \mbox{\cite{yu2012control,tsaturyan2017ultracoherent}}, and therefore suppressing curvature near the clamping regions is an effective strategy, as expected for soft-clamped resonators \mbox{\cite{tsaturyan2017ultracoherent}}. Several approaches have been developed to realize this idea, including tapered clamping \mbox{\cite{bereyhi2019clamp}}, hierarchical structuring \mbox{\cite{bereyhi2022hierarchical,fedorov2020fractal}}, phononic bandgap engineering \mbox{\cite{cupertino2024centimeter,beccari2022strained,ghadimi2018elastic}}, and perimeter modes in polygon resonators  \mbox{\cite{bereyhi2022perimeter}}.

Remarkably, the perimeter modes in polygon resonators have emerged as promising candidates for integrated quantum optomechanical systems \cite{bereyhi2022perimeter}. In the tapered-clamping approach \cite{bereyhi2019clamp}, enhancement of $Q$ is challenging, as it is constrained to the material’s yield stress-to-deposition stress ratio. In the hierarchical structuring approach \cite{bereyhi2022hierarchical,fedorov2020fractal}, the torsional losses in cascaded branchings hinder the resonators from attaining the "clampless" limit. Furthermore, in comparison to the phononic crystals (PnCs) \cite{beccari2022strained,cupertino2024centimeter,ghadimi2018elastic}, the fundamental perimeter modes of polygon structures \cite{bereyhi2022perimeter} have been shown to reach a quality factor of $3.6\times10^9$ in a compact device geometry. This compact architecture is advantageous for practical integration with optical microcavities, as it facilitates the maintenance of stable submicron gaps, a prerequisite for achieving strong near-field optomechanical coupling \cite{guo2019feedback}. Additionally, soft clamping in PnCs is typically realized in defect-localized bandgap modes. While bandgap engineering can provide excellent spectral isolation, practical finite-size devices may still host multiple localized defect modes and, depending on boundaries and fabrication disorder, occasional spurious resonances within or near the nominal bandgap. Importantly, soft-clamped perimeter modes in polygon resonators also exhibit broad frequency scalability with a quality factor $Q$ greater than $10^8$, maintained across more than four octaves ($170~ \text{kHz}-2.5~\text{MHz}$). Moreover, the $Q$s of the perimeter modes exhibit a high thermal-noise-limited force sensitivity of $420~\text{zN}\sqrt{\text{Hz}}$ at room temperature, which is comparable to that of atomic force microscopy cantilevers operating at millikelvin temperature ($190~\text{zN}\sqrt{\text{Hz}}$). By co-integrating low-loss perimeter-mode resonators with on-chip optical cavities, the combination of low mechanical dissipation and efficient near-field optomechanical readout can suppress both thermomechanical and readout noises, thereby facilitating the study of intrinsic frequency fluctuations. The resulting nonlinear optomechanical transduction yields signals with significantly reduced frequency noise, providing a suitable platform for precise frequency tracking \cite{kharbanda2026chip}.

While perimeter modes in polygon resonators provide an attractive platform for compact soft-clamped nanomechanical systems, Ref.~\cite{bereyhi2022perimeter} established an effective short-tether analytical description of perimeter modes and their dissipation dilution. Building on this description, we develop here a polygon-specific analytical model in which the torsional angle distribution along the supporting tethers is analytically derived from the modified Timoshenko–Gere (TG) equation. This formulation captures finite-wavelength torsional effects and extends the theory beyond the conventional short-tether limit. The resulting model reveals that dissipation dilution is strongly influenced by the coupling between polygon-side flexural motion and tether torsional dynamics. Torsional rotation at the tether–polygon junction modulates the tensile energy stored in the perimeter mode, while the supporting tethers simultaneously introduce an additional torsional dissipation channel. Importantly, the strength of this loss channel can be tailored by changing the tether length, thereby enabling optimization of the mechanical quality factor.

\begin{figure}[H]
  \centering
  \includegraphics[scale = 0.35]{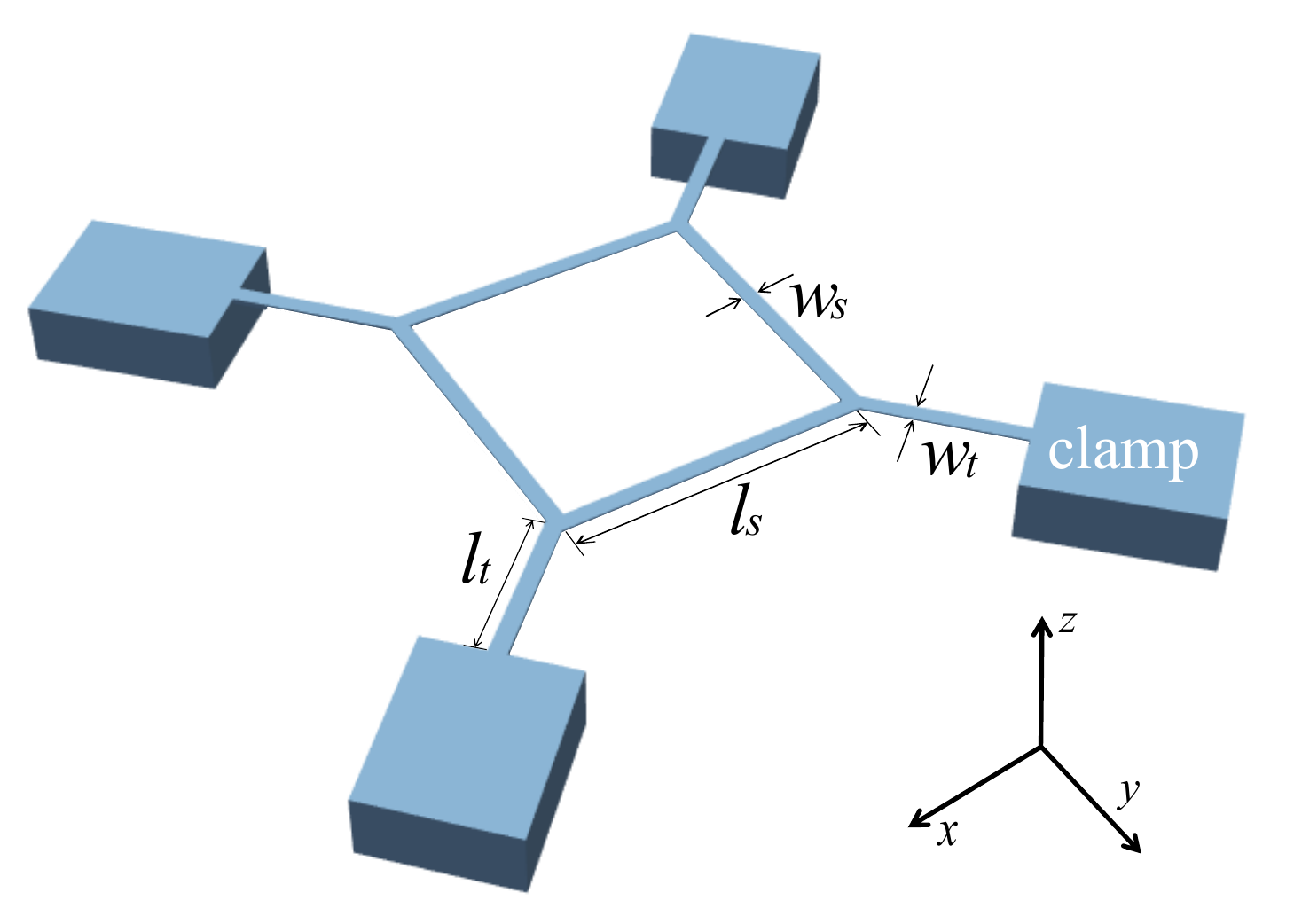}
  \caption{Schematic representation of the polygon resonator (light blue) with index number $N = 4$ and clamps (light blue). The substrates are represented in black. $l_{s}$, $l_{t}$, $w_{s}$, $w_{t}$ represent polygon side length, tether length, polygon side width, and tether width, respectively.}
\label{polygon-sketch}
\end{figure}

This article is structured as follows. First, we identify the two dominant dissipation channels in polygon resonators within the narrow-beam approximation. We then derive the flexural motion of the polygon sides and the torsional dynamics of the supporting tethers. By coupling these motions at the tether–polygon junctions, we obtain an analytical expression for the dissipation dilution factor. Next, we compare the analytical predictions with finite-element-method (FEM) simulations of representative silicon nitride (Si$_3$N$_4$) polygon resonators. Finally, we examine the validity and limitations of the analytical model across a broad geometric parameter space and identify the regimes in which neglected polygon-side torsional and warping deformations become significant.

\paragraph{Loss analysis. ---}  
\label{sec:develop}
Dissipation dilution can be fundamentally understood through the interaction between a harmonic oscillator and a lossless potential \cite{gonzalez2000suspensionsthermal}. The quality factor of a mechanical mode subject to dissipation dilution is given by 

\begin{equation}
Q = D_Q \cdot Q_{\text{int}}. 
\end{equation}

Here, $D_Q$ is the dissipation dilution factor, and $Q_{\text{int}}$ is the material's intrinsic quality factor, which is inversely proportional to the material loss angle ($Q_{\text{int}} = 1/\phi$) \cite{saulson1990thermal}. For the $n^{\text{th}}$ resonator mode, the dissipation dilution factor $D_Q$ can be written in the general form as \cite{Unterreithmeier2009,fedorov2019generalized,fedorov2020fractal}:

\begin{equation}
D_{\text{Q},n} =\frac{\langle W_{\text{total}} \rangle}{\langle W_{\text{lossy}} \rangle}= \frac{1}{\alpha_n\lambda + \beta_n\lambda^2},
\label{eq:dilution}
\end{equation}

\noindent 
where $\langle W_{\text{total}}\rangle$ denotes the total stored energy, and $\langle W_{\text{lossy}}\rangle$ is the lossy part. The coefficients $\alpha_n$ and $\beta_n$ are determined by the resonator's boundary curvature and the distributed bending over the rest of the mode, respectively. The strain parameter $\lambda = \frac{h}{l}\sqrt{\frac{E}{12\sigma}}$ depends on the resonator length $l$, thickness $h$, Young's modulus $E$, and tensile stress $\sigma$.

\begin{figure}[H]
  \centering 
  \includegraphics[scale = 0.6]{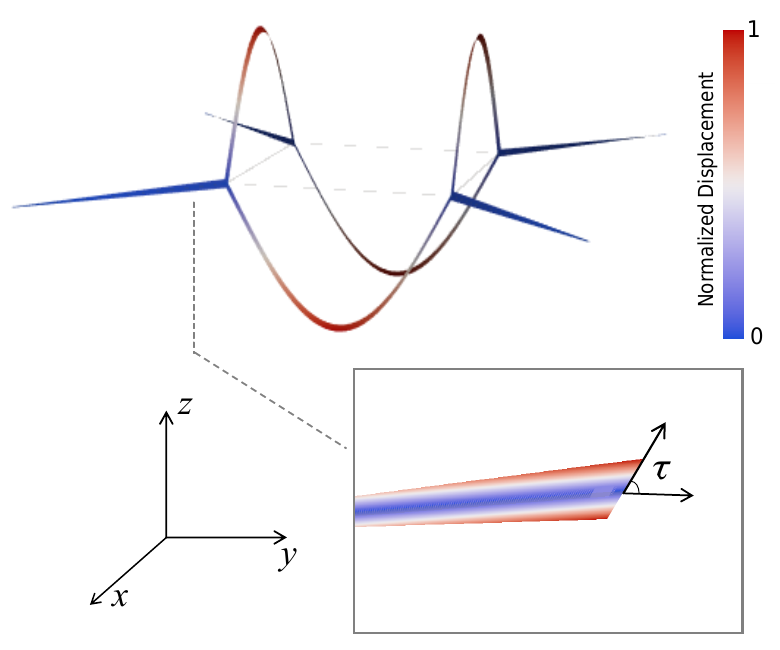}
  \caption{FEM simulation of the fundamental mode of a stress-preserving polygon resonator. The inset illustrates the torsion deformation caused by the out-of-plane displacement of two neighboring polygon sides. $\tau$ is the torsion angle of the supporting tethers.}
  \centering
\label{displacement-3}
\end{figure}

To study soft-clamped modes in polygon resonators, we start with a four-interconnected beam structure as a general example, as illustrated in \figref{polygon-sketch}. Each beam is supported by two tethers, located at the vertices of the polygon. Due to the underlying symmetry, the fundamental perimeter mode possesses a symmetric displacement profile, as shown in \figref{displacement-3}. The transverse reaction forces on opposing tethers demonstrate phase cancellation, resulting in negligible net bending moments. However, the out-of-plane tangential displacement components generate a non-zero torque about the tether axes, leading to torsional motion and the associated dissipation discussed below.

In particular, as a result of practical design considerations, all constituent elements are assumed to operate in the narrow-beam regime, with lengths much greater than their widths. Under this approximation, perimeter modes do not induce flexural displacements in the supporting tethers, and the corresponding boundary-loss coefficient vanishes ($\alpha_n=0$) \cite{bereyhi2022perimeter}. Therefore, the dominant dissipation pathways in polygon resonators arise from (i) bending of the polygon sides and (ii) torsional strain energy in the supporting tethers. To analytically estimate the dissipation dilution factor of the perimeter modes, we first derive the equations governing the transverse displacement $u$ of the polygon sides and the torsional angle $\tau$ of the supporting tethers, respectively. We then incorporate the coupling between these motions into the analytical model.

\paragraph{Flexural dynamics of polygon sides. ---} To model the bending-dominated dissipation, we analyze the transverse displacement field $u(x)$ along each side of the $N$-sided polygon. Within the Euler-Bernoulli beam theory, $u(x)$ satisfies the flexural string equation in the high tension limit \cite{schmid2016fundamentals}:

\begin{equation}
-\sigma_s\frac{d^2 u}{d x^2} = \rho \omega_n^2 u.
\label{polygon side1}
\end{equation}

\noindent where $x$ is the axial coordinate along the polygon sides, $\sigma_s$ is the stress, and $\rho$ is the material density. Specifically, $\omega_n$ denotes the eigenfrequency of the $n^{\text{th}}$ flexural mode of an individual polygon side. Owing to the discrete rotational symmetry of the polygon, the corresponding perimeter modes are degenerate, sharing the same frequency while differing only in the phase relations between adjacent sides. The stress distribution $\sigma_{s}$ is determined by force balance and the constancy of the path-length between the two clamped vertices of the $N$-sided polygon \cite{bereyhi2022perimeter}:

\begin{equation}
    \sigma_{s} = \frac{\sigma_0 (1-\nu)[1+ 2r_{l}\sin{(\pi/N)}]}{1 + 4(r_{l}/r_{w})\sin^2{}(\pi/N)},
\end{equation}

\noindent with $\sigma_0$ representing the initial deposition stress, $\nu$ the Poisson's ratio, $r_{l} = l_{s}/l_{t}$ the ratio of the polygon side length to the tether length, and $r_{w} = w_{s}/w_{t}$ the ratio of the polygon side width to the tether width. Under the narrow-beam and high-tension conditions ($\lambda \ll 1$), the transverse motion of each polygon side is tension dominated. Consequently, the transverse displacement can be well approximated by the sinusoidal mode shape $u(x) = u_0\sin{(n\pi x/l_s)}$ with polygon side length $l_s$. $\omega_{n}$ can be determined by substituting $u(x)$ into \eqref{polygon side1}:

\begin{equation}
     \omega_n^2 = \frac{\sigma_{s}}{\rho}\frac{n^2 \pi^2}{l_{s}^2}.
\end{equation}

\paragraph{Torsional vibration of supporting tethers. ---}

The torsion angle $\tau$ is determined by the modified TG equation, incorporating the axial stress term (see Appendix A for the full derivation of the modified TG equation):

\begin{equation}
    EI_{\psi} \frac{\partial^4 \tau}{\partial x^4} +(\rho I_{\psi}\omega_n^2-C - \sigma_t I_y)\frac{\partial^2 \tau}{\partial x^2}-\rho I_p \omega_n^2 \tau =0.
    \label{TG equation}
\end{equation}
\noindent where $\sigma_t = \sigma_{xx}$ denotes the axial stress along the tether, $x,y,z$ are the Cartesian coordinates, $I_p = \iint_A (y^2 +z^2)\mathrm{d}A$ is the geometric moment of inertia, $I_{\psi} = \iint_A \psi^2 \mathrm{d}A$ ($\psi$ $\approx -yz$ the warping function \cite{chopin2019extreme}), $A$ the cross section with thickness $h$ and tether width $w_t$, $C$ the torsional rigidity ($C = \iint_A G [(\frac{\partial \psi}{\partial y} -z)^2 + (\frac{\partial \psi}{\partial z}+y)^2]\mathrm{d}A$), and $G$ the shear modulus. Physically, torsional deformation of a highly stressed beam causes the axial pre-stress to provide an additional restoring torque through geometric stiffening. This effect contributes a stress-dependent torsional restoring term proportional to $\sigma_t I_y$, increasing the effective torsional stiffness to $C+\sigma_t I_y$. Under the narrow-beam approximation, the warping-related fourth-order term ($EI_\psi \tau''''$) and the associated warping-inertia correction ($\rho I_\psi \omega_n^2 \tau''$) are neglected; thus, \eqref{TG equation} reduces to

\begin{equation}
    -(C+ \sigma_{t}I_{y})\frac{d^2\tau}{dx^2} = \rho I_{\text{p}} \omega_n^2\tau,
    \label{tether1}
\end{equation}

 \noindent where $C+ \sigma_{t}I_{y}$ is the effective torsional rigidity enhanced by the axial stress. With the clamped-end boundary condition $\tau(0)=0$, the torsion profile is written as $\tau(x)=\tau_0\sin(kx)$. Substituting this expression into \eqref{tether1} and using the frequency $\omega_n$ of the polygon side gives

\begin{equation}
    k = \sqrt{I_{\text{p}}\sigma_{s}\frac{n^2\pi^2}{l_{s}^2}\frac{1}{C+\sigma_{t}I_y}}.
    \label{wavenumber}
\end{equation}

\paragraph{Coupling at the branch nodes. ---}
The flexural deformation of the polygon sides generates a torque that is directly coupled to the supporting tethers, initiating torsional wave propagation along the tether elements. The coupling between the transverse and torsional vibrations can be characterized by the following relation \cite{fedorov2020fractal}:

\begin{equation}
     u^{\prime}(0) = \sin[\frac{(N-2)\pi}{2N}] \tau(l_{t}) = \cos{(\frac{\pi}{N})}\tau_0\sin{(kl_{t})}.
     \label{u'formula}
\end{equation}

\eqref{u'formula} shows that the modal gradient at the branch nodes, $u^{\prime}(0)$, is controlled by the geometric factor $\cos(\pi/N)$ and the accumulated torsional phase through $\sin(kl_t)$. Because the tether-torsion contribution is controlled by the modal slope at the polygon-tether junction, \eqref{u'formula} provides two simple design guidelines within the model: increasing the geometric factor $\cos{(\pi/N)}$ and engineering the accumulated torsional phase through the tether length. For the tether-torsion contribution alone, the optimum condition occurs near $\sin{(kl_t)}= 1$. This relation also clarifies the role of the tether length. Within the analytical model, when the material properties and the remaining geometric parameters are fixed so that $k$ is approximately constant, varying $l_t$ directly tunes the accumulated torsional phase $kl_t$, thereby controlling the flexural–torsional coupling between the polygon sides and the supporting tethers.

\paragraph{Dissipation dilution factors and loss coefficient. ---}
The dissipation dilution factor $D_{Q}$ of perimeter modes can be determined by estimating different energy components in \eqref{eq:dilution}. The "lossless" tensile energy is given by \cite{fedorov2020fractal}

\begin{equation}
     \langle W_{\text{tens}}\rangle = \frac{N}{2}\sigma_{s}w_{s}h\int_{0}^{l_{s}}[u^{\prime}(x)]^2 dx 
     \label{Wtens1}.
\end{equation}

The lossy energy comprises two components. The distributed bending energy along the polygon sides is 

\begin{equation}
    \langle W_{\text{bend}}\rangle = \frac{N}{2}\frac{Ew_{s}h^3}{12} \int_{0}^{l_{s}}[u^{\prime\prime}(x)]^2 dx,
    \label{Wbend1}
\end{equation}

\noindent and the torsional lossy part is

\begin{equation}
    \langle W_{\text{tors}}\rangle = N\frac{Gw_{t}h^3}{12} \int_{0}^{l_{t}}(\tau^{\prime}(x))^2  dx.
    \label{Wtors1}
\end{equation}

Substituting Eqs.~\ref{Wtens1}--\ref{Wtors1} into \eqref{eq:dilution} and defining $\lambda_l = \frac{h}{l_{s}}\sqrt{\frac{E}{12\sigma_{s}}}$ yields 

\begin{align}
\begin{split}
    D_{Q,n}^{-1}& =\frac{\langle W_{\text{bend}}\rangle+\langle W_{\text{tors}}\rangle}{\langle W_{\text{tens}}\rangle}
    \\
    &=D_{Q,\text{bend}}^{-1} + D_{Q,\text{tors}}^{-1}.
\end{split}
\label{DQgeneral1} 
\end{align}
\noindent with the two lossy contributions
\begin{equation}
    D^{-1}_{Q,\text{bend}} = n^2\pi^2\lambda_l^2,
    \label{DQbend1}
\end{equation}
\noindent and
\begin{equation}
    D^{-1}_{Q,\text{tors}} = \lambda_l^2 \frac{kl_{s}}{(1+\nu)r_{w}}\frac{2kl_{t}+\sin{2kl_{t}}}{\cos^2(\pi/N)\sin^2(kl_{t})}.
    \label{DQtors1}
\end{equation}

\begin{figure*}[t]
  \centering 
  \includegraphics[scale = 0.9]{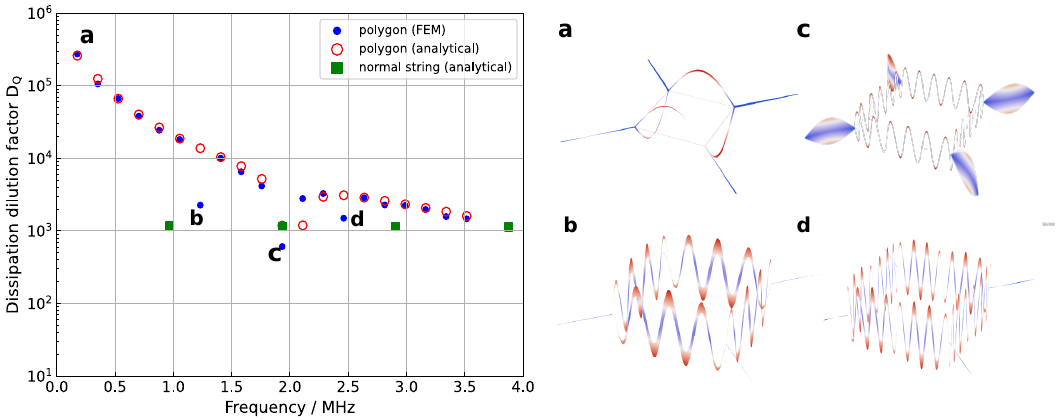}
\caption{\justifying
\hyphenpenalty=25
Dissipation dilution factors and frequencies of the perimeter modes of the resonator with dimensions shown in the simulation section. Blue filled circles and red open circles correspond to FEM results and analytical results, respectively. Green squares correspond to out-of-plane modes of normal strings with the same dimensions and mechanical properties as the polygon sides. The labels a--d next to selected FEM data points in the left panel identify the corresponding mode shapes shown on the right. Mode a is the fundamental perimeter mode, while modes b--d are higher-order perimeter modes exhibiting mode hybridization.}
\label{Qf}
\end{figure*}

The first term in \eqref{DQgeneral1} is due to the distributed bending across the polygon sides, and sets an upper bound on the dissipation dilution factor. The second term is related to the torsional deformation of the supporting tethers. Notably, the factor $\sin^2{(kl_{t})}$ in \eqref{DQtors1} shows that the dissipation dilution factor is strongly influenced by the accumulated torsional phase in the tethers. As $kl_t$ approaches an integer multiple of $\pi$, the torsional contribution to the lossy energy increases substantially, and leads to a pronounced reduction of the dilution factor. Conversely, favorable dilution is obtained near $kl_t=(2m+1)\pi/2$, with the first maximum generally being the most relevant regime. Eqs.~\ref{DQbend1} and \ref{DQtors1} predict the characteristic $1/\lambda^2$ scaling of the dissipation dilution factor, consistent with Ref.~\cite{bereyhi2022perimeter}.

\paragraph{Simulation. ---}
As an illustrative case, we consider a polygon resonator characterized by the following geometric parameters: vertex count $N=4$, side length $l_{s}=1400~\upmu\mathrm{m}$, polygon side width $w_{s}=300~\mathrm{nm}$, support length $l_{t}=700~\upmu\mathrm{m}$, and support width $w_{t}=400~\mathrm{nm}$. The resonator structure is made of a $100~\mathrm{nm}$ commercial Si$_3$N$_4$ thin film with Young's modulus $E = 250~\mathrm{GPa}$, mass density $\rho = 3100~\mathrm{kg/m^3}$, Poisson's ratio $\nu = 0.23$, and deposition stress $\sigma_0 = 1~\mathrm{GPa}$.

\begin{figure*}[t]
    \centering
    \includegraphics[scale = 0.42]{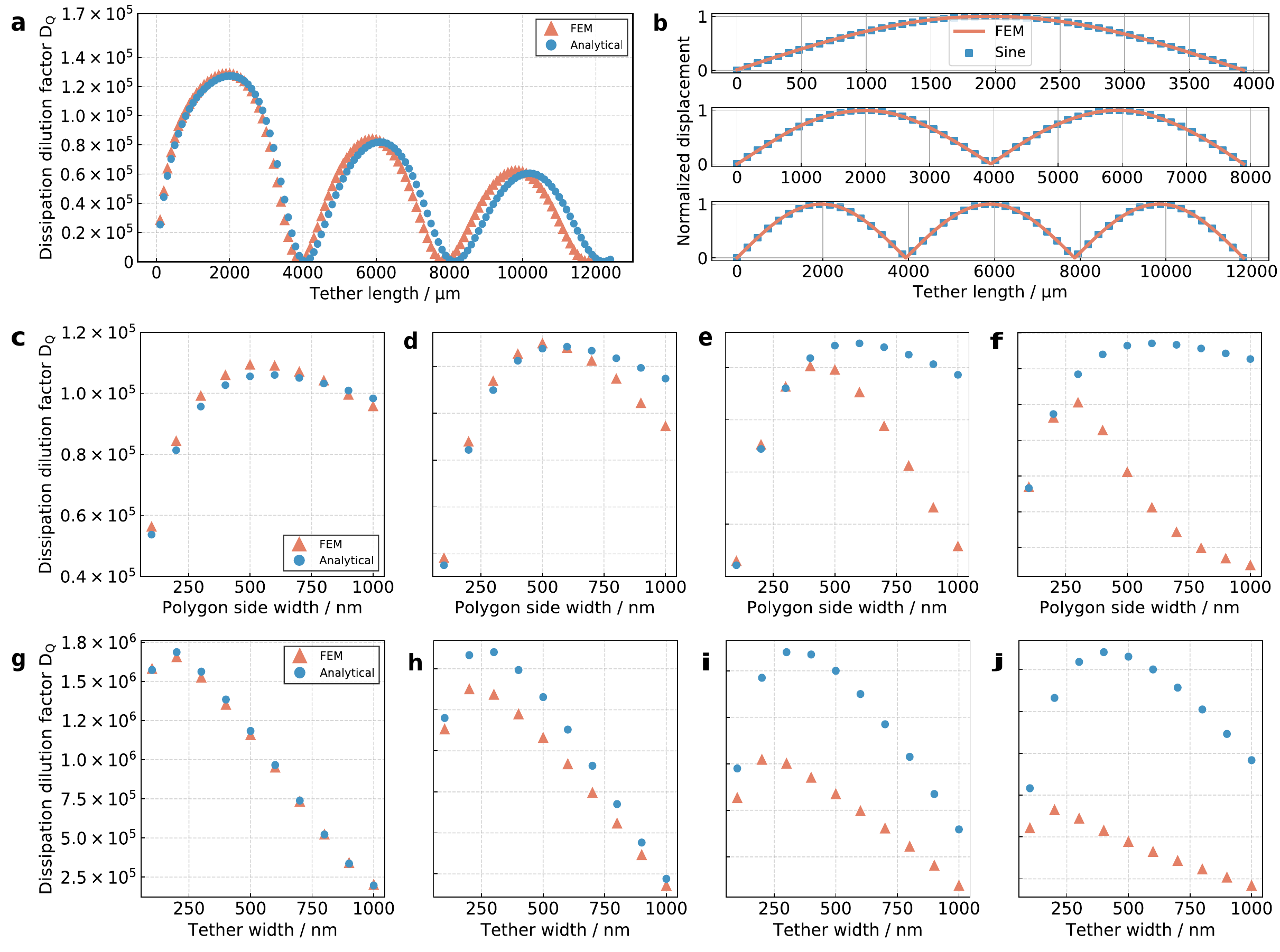}
     \caption{\justifying  
        \hyphenpenalty=25  
     Dissipation dilution factors of fundamental perimeter modes with $N=4$. Analytical predictions are shown in light blue, and FEM simulations in orange. (a) $D_Q$ as a function of tether length for $l_s=700~\upmu\mathrm{m}$, $w_s=300~\mathrm{nm}$, and $w_t=400~\mathrm{nm}$. (b) Torsion-angle profiles of the tethers for lengths of 3920, 7890, and $11810~\upmu\mathrm{m}$. (c–f) $D_Q$ as a function of polygon-side width for device thicknesses of 80, 60, 40, and $20~\mathrm{nm}$, respectively, with $l_s=700~\upmu\mathrm{m}$ and $l_t=350~\upmu\mathrm{m}$. (g–j) $D_Q$ as a function of tether width for polygon-side widths of 200, 400, 600, and $800~\mathrm{nm}$, respectively, with the thickness fixed at $20~\mathrm{nm}$.}
\phantomsubcaption\label{fig4a} 
    \phantomsubcaption\label{fig4b} 
    \phantomsubcaption\label{fig4c} 
    \phantomsubcaption\label{fig4d} 
    \phantomsubcaption\label{fig4e} 
    \phantomsubcaption\label{fig4f} 
    \phantomsubcaption\label{fig4g} 
    \phantomsubcaption\label{fig4h} 
    \phantomsubcaption\label{fig4i} 
    \phantomsubcaption\label{fig4j} 

    \label{fig4_combined}
\end{figure*}

To evaluate the accuracy of the analytical model, we present a comparative analysis of resonance frequencies and dissipation dilution factors for 20 modes of the polygon resonator in \figref{Qf}, as obtained from FEM simulations and the analytical model. The analytical model predicts the resonance frequencies with relative errors less than 0.1\%. For most modes, the analytically estimated dilution factors are in good quantitative agreement with the FEM results. However, deviations are observed for the $7^{\mathrm{th}}$ and $14^{\mathrm{th}}$ perimeter modes, corresponding to the FEM data points labeled b and d in \figref{Qf}. The mode shapes associated with these two points reveal that the polygon-side motion is no longer purely flexural, but contains a significant torsional component. This flexural--torsional hybridization redistributes part of the strain energy into torsional deformation, thereby reducing the dissipation dilution factor. In addition, the $11^{\mathrm{th}}$ and $12^{\mathrm{th}}$ modes, corresponding to the FEM data point labeled c in \figref{Qf}, also deviate from the analytical predictions because the tether motion is no longer purely torsional, but exhibits simultaneous torsional and flexural deformation. This bending--torsion hybridization introduces additional strain-energy contributions that are not captured by the analytical model. Remarkably, the dilution factor of the fundamental mode, corresponding to the FEM data point labeled a in \figref{Qf}, is approximately two orders of magnitude higher than that of a simple doubly-clamped beam whose dimensions match those of a single polygon side.

To investigate the coupling between the flexural modes of the polygon sides and the torsional modes of the tethers, we fix $l_{\text{s}} = 700~\upmu \mathrm{m}$ and vary the tether length while keeping all other geometric parameters constant. \figref{fig4a} plots the dissipation dilution factor as a function of tether length, revealing several local maxima associated with torsional phase accumulation in the tethers. The first maximum gives the largest dissipation dilution, while subsequent maxima are reduced as an increasing fraction of the total elastic energy is stored in torsional deformation of the tethers. This behavior originates from the torsional rotation of the tether at the polygon–tether junction, which modifies the boundary condition of the polygon-side motion. Through the coupling between polygon-side bending and tether torsion, the accumulated torsional phase alters the partition of elastic energy between tensile and torsional contributions, thereby influencing the dissipation dilution factor. Consistent with the model, the first maximum near $k l_t \approx \pi/2$ and the first minimum near $k l_t \approx \pi$ agree well with \eqref{DQtors1}. The torsion-angle profiles near successive phase nodes, $k l_t\approx m\pi$ with $m=1,2,3$, corresponding to tether lengths of 3920, 7890, and $11810~\upmu \mathrm{m}$, are further visualized in \figref{fig4b}. These results demonstrate that tether length serves as an effective design parameter for controlling the dissipation dilution factor through torsional phase accumulation. As the tether length increases, the peak positions predicted by the analytical model exhibit a growing deviation from the FEM simulations. One possible contribution to this discrepancy is the length dependence of the tensile stress, which has been observed experimentally in highly stressed nanomechanical string resonators \mbox{\cite{buckle2021universal}}. Since the analytical model assumes a prescribed stress value, variations of the actual stress with tether length can modify the torsional wave number and consequently shift the phase condition.

We now examine the applicability of the analytical model by studying how its accuracy changes with the polygon-side width. This comparison identifies the regime in which the quasi-one-dimensional approximation remains quantitatively reliable. In \figref{fig4c}-\ref{fig4f}, the polygon side length and tether length are fixed at $700~\upmu\mathrm{m}$ and $350~\upmu\mathrm{m}$, respectively, while the polygon-side width is swept from 100 to $1000~\mathrm{nm}$ for four different thicknesses, $h=80$, 60, 40, and $20~\mathrm{nm}$. For $h=80~\mathrm{nm}$, the analytical model agrees well with the FEM results over the entire width range. As the thickness decreases, however, noticeable deviations appear when the polygon-side width exceeds approximately 600, 400, and $200~\mathrm{nm}$ for $h=60$, 40, and $20~\mathrm{nm}$, respectively. These critical widths correspond to a nearly constant aspect ratio of $w_s/h\approx10$, indicating that the validity of the analytical model is mainly limited by the polygon-side width-to-thickness ratio.

We then examine the effect of tether width by fixing the thickness at $20~\mathrm{nm}$ and sweeping the tether width from 100 to $1000~\mathrm{nm}$ in \figref{fig4g}-\ref{fig4j}. The polygon-side widths are fixed at 200, 400, 600, and $800~\mathrm{nm}$, respectively. For $w_s=200~\mathrm{nm}$, corresponding to $w_s/h\approx10$, the analytical model remains in good agreement with the FEM results throughout the tether-width range. When $w_s$ is increased to 400, 600, and $800~\mathrm{nm}$, the discrepancy between the analytical prediction and FEM becomes increasingly pronounced, although the tether width is varied over the same range. This comparison indicates that the deviation from the analytical model is governed mainly by the polygon-side geometry rather than by the tether width.

The comparison of these two scans shows that the analytical model remains quantitatively reliable below a critical polygon-side width-to-thickness ratio and deviates once this ratio becomes too large. Since the discrepancy follows $w_s$ much more strongly than $w_t$, it is likely associated with deformation of the polygon sides that is neglected in the present model. For large $w_s/h$, the polygon side can no longer be treated as an ideal one-dimensional, tension-dominated beam. Torsional and/or warping deformation of the polygon side may then add lossy energy contributions not included in the analytical estimate, leading to the observed reduction in dissipation dilution.

\paragraph{Torsional dissipation dilution limit. ---}
Finally, we discuss a separate limitation associated with torsional dissipation dilution in the supporting tethers. A related problem has been analyzed for doubly-clamped stressed nanoribbons, whose geometry closely resembles that of the supporting tethers (see Appendix B for a detailed derivation). Applying this estimate to the tethers of a polygon resonator gives the dissipation dilution factor for the $n^{\text{th}}$ torsional mode:

 \begin{equation}
    D_{Q,t,n} \approx 1 +\frac{1}{\frac{2G}{\sigma_t}\frac{h^2}{w_t^2}+\frac{n^2\pi^2}{12}\frac{Eh^2}{\sigma_t l_t^2} + \frac{h}{l_t}\sqrt{\frac{E}{3\sigma_t}}}.
\label{DQrib}
\end{equation}
  
The dimensions of the tethers ($l_t \gg w_t$) and material properties of the Si$_3$N$_4$ film (see the simulation section) suggest that the first term dominates the denominator in \eqref{DQrib}. This finding aligns with prior work \cite{pratt2023nanoscale} and points to a limitation in torsional dissipation dilution.
  
\paragraph{Conclusions and outlook. ---}
In this work, we present an analytical model of perimeter vibrational modes in polygonal resonators. We derive the TG equation for the angular displacement of nanoribbons under high tension. The resulting torsional equation is applied to the supporting tethers of polygon resonators and combined with the flexural motion of the polygon sides to estimate the resonance frequencies and dissipation dilution factors of perimeter modes. Notably, the quality factor of the fundamental mode is enhanced by two orders of magnitude compared to uniform strings of the same dimensions as the polygon sides. We show that dissipation dilution in perimeter modes is governed by the coupling between polygon-side bending and tether torsion, with the tether length serving as an effective tuning parameter through the accumulated torsional phase. The resulting model reproduces the characteristic $1/\lambda^2$ scaling of stressed polygon resonators and clarifies its physical origin beyond the conventional short-tether approximation. These results provide practical design guidelines for achieving high mechanical quality factors. We further establish the torsional dissipation dilution limit of the supporting tethers, with potential applications in nanoscale force sensing \mbox{\cite{moser2013ultrasensitive,gavartin2012hybrid}}. More broadly, the demonstrated control of bending–torsion interactions may be useful for studying multimode dynamics, non-Hermitian physics, and topological wave phenomena in nanomechanical systems \mbox{\cite{faust2012nonadiabatic,peano2015topological,lu2017observation}}.

\section{Acknowledgements}
 Y. Z. acknowledges the support from the National Natural Science Foundation of
China (Grant No. 12474153, Grant No. 11804163), Guangdong Provincial Quantum Science Strategic Initiative (GDZX2401001).

\appendix
\renewcommand{\theequation}{A\arabic{equation}}  
\setcounter{equation}{0}  
\section{Appendix A: Free Torsional Vibration Under Tensile Stress} 
\label{appendix1}

The theoretical foundation of torsional dissipation dilution can be traced back to Buckley's pioneering work on twisted elastic strips \cite{buckley1914lxxxiv}, which demonstrates that tensile stress can geometrically enhance torsional stiffness without proportionally increasing dissipative losses. This phenomenon arises because tensile stress preferentially aligns the principal strain axes along the ribbon's length, thereby suppressing transverse deformation through geometric stiffening, while intrinsic loss mechanisms remain primarily governed by material properties \cite{trahair2017flexural} (p258-265). Buckley's analysis of bifilar systems reveals that torsion modes of ribbons are naturally soft-clamped \cite{tsaturyan2017ultracoherent} and the effective quality factor $Q_{\mathrm{eff}}$ can be enhanced beyond the material limit through stress engineering, a critical insight for modern nanomechanical resonators \cite{fedorov2019generalized}.

The study by Pratt et al. \cite{pratt2023nanoscale} significantly extends Buckley's paradigm by demonstrating experimentally the validation of extreme dissipation dilution ($Q\cdot f > 10^{13}$ Hz) in nanoribbons made of $\text{Si}_3\text{N}_4$ at room temperatures. Their work establishes a novel theoretical framework for quantum precision measurements exploiting the torsion modes of nanomechanical resonators \cite{kim2016approaching,enomoto2016standard,treps2003quantum}, enabling novel designs where torsional modes benefit from dissipation dilution \cite{bereyhi2022perimeter,ghadimi2018elastic}. 

Here, we derive the general solution for the torsional vibrational modes of a stressed nanoribbon based on Hamilton’s variational principle. First, we revisit the TG theory for torsional vibration of ribbons in the cartesian representation \cite{rao2019vibration} (p296-302). For a normal nanoribbon resonator with length $L$, width $w$, and thickness $h$ satisfying $L > w > h$, the equation of motion for the angular displacement $\Theta(x,t)$ is

\begin{align}
\begin{split}
    \rho I_p \frac{\partial^2 \Theta}{\partial t^2} - \frac{\partial^2}{\partial x \partial t}(\rho  I_{\psi}\frac{\partial^2 \Theta}{\partial x \partial t}) &- \frac{\partial}{\partial x}(C \frac{\partial \Theta}{\partial x}) 
    \\
    &+ \frac{\partial^2}{\partial x^2}(E I_{\psi}\frac{\partial^2 \Theta}{\partial x^2}) = 0,
    \label{e1}
\end{split}
\end{align}

\noindent where $x$ denotes the axial coordinate along the ribbon, $I_p = \iint_A (y^2 +z^2)\mathrm{d}A$ is the geometric moment of inertia, $I_{\psi} = \iint_A \psi^2 \mathrm{d}A$ ($\psi$ $\approx -yz$ is the warping function \cite{chopin2019extreme}), $A = w\cdot h$ the cross section, $C$ the torsional rigidity ($C = \iint_A G [(\frac{\partial \psi}{\partial y} -z)^2 + (\frac{\partial \psi}{\partial z}+y)^2]\mathrm{d}A$), $E$ the Young's modulus, $G$ the shear modulus, and $\rho$ the material density. In the case of a highly stressed ribbon, the angular displacement is obtained by minimizing the action integral $\mathcal{S} = \int_{t_1}^{t_2} (T - W_{\text{strain}}-W_{\text{tens}}) \, dt$, where $T$, $W_{\text{strain}}$ and $W_{\text{tens}}$ represent the kinetic, elastic strain and tensile energies, respectively. The modified TG equation can be written as 

\begin{align}
\begin{split}
    \rho I_p \frac{\partial^2 \Theta}{\partial t^2} - \frac{\partial^2}{\partial x \partial t}(\rho  I_{\psi}\frac{\partial^2 \Theta}{\partial x \partial t}) &- \frac{\partial}{\partial x}((C + \sigma I_y)\frac{\partial \Theta}{\partial x}) 
    \\
    &+ \frac{\partial^2}{\partial x^2}(E I_{\psi}\frac{\partial^2 \Theta}{\partial x^2}) = 0,
    \label{e11}
\end{split}
\end{align}

\noindent where $I_{y}$ is the geometric moment of inertia with respect to the $y$-axis, and $\sigma = \sigma_{xx}$ is the axial stress component. Physically, torsional motion of a highly stressed ribbon slightly elongates the ribbon centerline and therefore couples the twist dynamics to the axial stress field. This geometric stiffening produces an additional restoring term proportional to $\sigma I_y$, so that the effective torsional rigidity becomes $C+\sigma I_y$ in the high-stress limit. By applying separation of variables, the solution to this differential equation takes the form of a superposition of normal modes, each consisting of a position-dependent component multiplied by a time-dependent component:

\begin{equation}
    \Theta(x,t) = \sum_{n = 1}^{\infty} u_{0,n}\cos{(\omega_n t)}\theta_n,
    \label{Ux}
\end{equation}

\noindent where $\omega_n$ is the eigenfrequency of the $n^{\text{th}}$ mode, $\theta_n$ the mode shape, and $u_{0,n}$ the amplitude. Inserting \eqref{Ux} into \eqref{e11} yields

\begin{equation}
    EI_{\psi} \frac{\partial^4 \theta_n}{\partial x^4} +(\rho I_{\psi}\omega_n^2-C - \sigma I_y)\frac{\partial^2 \theta_n}{\partial x^2}-\rho I_p \omega_n^2 \theta_n =0.
    \label{e5}
\end{equation}

\noindent Assuming a sinusoidal mode shape with wavenumber $\beta_n$, and applying the boundary conditions $\theta_n(0) = \theta_n(L) = \frac{\partial^2 \theta_n}{\partial x^2}\vert_0 = \frac{\partial^2 \theta_n}{\partial x^2}\vert_L = 0$ yield

\begin{equation}
    \theta_n(x) = \sin{(\beta_n x)}, ~\beta_n = \frac{n\pi}{L}.
    \label{e66}
\end{equation}

\noindent Combining \eqref{e66} with \eqref{e5} gives

\begin{align}
\begin{split}
    \rho w^2h^2\beta_n^2\omega_n^2 + 12\rho(w^2 +h^2)\omega_n^2 &= Ew^2h^2\beta_n^4 
    \\
    &+ 48Gh^2\beta_n^2 +12w^2\sigma \beta_n^2.
\end{split}
\end{align}

The first term $\rho w^2 h^2 \beta_n^2 \omega_n^2$ is negligible due to $\rho w^2 h^2 \beta_n^2  \ll 12\rho(w^2 + h^2)$. Similarly, the term $Ew^2h^2\beta_n^4$ exhibits negligible influence on the system dynamics compared to the two dominant terms dominating the right-hand side. The mode frequency $\omega_n$ can be written in the following form

\begin{align}
\begin{split}
    \omega_n &= \sqrt{\frac{ 48Gh^2\beta_n^2 + 12w^2\sigma \beta_n^2}{ 12\rho (w^2 + h^2)}}
    \\
    & \approx \frac{n\pi}{L}\sqrt{\frac{\sigma}{\rho}(1 + \frac{4G}{ \sigma}\frac{h^2}{w^2})}.
\end{split}
    \label{e6}
\end{align}

\noindent The full set of wave numbers can be found by solving the dispersion relation of the pre-stressed ribbon:

\begin{equation}
    \beta_n^4 + (\frac{48G}{Ew^2} + \frac{12\sigma}{Eh^2})\beta_n^2 -\frac{12\rho}{Eh^2}\omega_n^2 = 0.
\end{equation}

\noindent Four solutions of $\beta_n$ can be found:

\begin{equation}
    \beta_{n,1-2} =\beta_{\sigma}= \frac{n\pi}{L},
\end{equation}

\begin{equation}
    \beta_{n,3-4} = \beta_{E} = \sqrt{\frac{12\sigma}{Eh^2}(1 + \frac{4G}{\sigma}\frac{h^2}{w^2})}.
\end{equation}

Within the analytical model, $\beta_{\sigma}$ is the already known wavenumber of a perfect ribbon, and $\beta_{E}$ is the wave number related to the torsional stiffness of the ribbon. Under clamped boundary conditions, the mode shape takes the following form:

\begin{equation}
    \theta_n(x) = \sin{(\beta_{\sigma}x)} + \frac{\beta_{\sigma}}{\beta_{E}}[e^{-\beta_{E}x} - \cos{(\beta_{\sigma}x)}].
\label{tor-theta}
\end{equation}

This equation enables the decoupling of the idealized torsion mode from the boundary-constrained components in the actual vibrational mode. It reveals that $\beta_E$ corresponds to the inverse characteristic length of the exponentially decaying edge effect.

\renewcommand{\theequation}{B\arabic{equation}}
\setcounter{equation}{0}  
\section{Appendix B: Torsional Dissipation Dilution of Nanoribbons} 
\label{appendix2}

The quality factor of a stressed nanomechanical resonator exhibits significant enhancement due to dissipation dilution. For a ribbon resonator under tensile stress $\sigma$, the dissipation dilution factor $D_{Q}$ can be written as

\begin{equation}
{D_{Q,n} = \frac{Q}{Q_{\text{int}}} = 1 + \frac{\langle W_{\text{tens}}\rangle}{\langle W_{\text{lossy}}\rangle}},
\label{DQn}
\end{equation}

\noindent where $\langle W_{\text{tens}}\rangle$ is the energy stored in the stress potential, and $\langle W_{\text{lossy}}\rangle$ is the energy concentrated in the bending and shear components \cite{fedorov2019generalized,sementilli2022nanomechanical}. The dissipation dilution factor $D_{Q}$ measures the proportional enhancement in the mechanical quality factor resulting from tensile stress. To analytically estimate $D_{Q}$, we express a lossless tensile energy based on \eqref{Ux}:

\begin{equation}
    \langle W_{\text{tens}}\rangle = \frac{\sigma}{2}\frac{hw^3}{12}\int u_{0,n}^2\langle(\frac{\partial \theta_n}{\partial x})^2\rangle dx.
    \label{tens}
\end{equation}

\noindent The lossy energy is

\begin{equation}
    \langle W_{\text{lossy}}\rangle = \frac{E}{2}\frac{wh^3}{12}\int u_{0,n}^2\langle \frac{2}{1+\nu}(\frac{\partial \theta_n}{\partial x})^2 + \frac{w^2}{12}(\frac{\partial^2 \theta_n}{\partial x^2})^2\rangle dx.
    \label{lossy}
\end{equation}

\noindent Substituting \eqref{tens} and \eqref{lossy} into \eqref{DQn} yields a dissipation dilution factor of 

\begin{equation}
    D_{\text{Q},n} = 1 + \frac{1}{\frac{2G}{\sigma}\frac{h^2}{w^2}+ \frac{n^2 \pi^2}{12}\frac{Eh^2}{\sigma L^2} + \frac{h}{L}\sqrt{\frac{E}{3\sigma}(1+\frac{4G}{\sigma}\frac{h^2}{w^2})}}.
    \label{DQ,n}
\end{equation}

The first two terms in the denominator arise from distributed shear deformation and distributed bending, respectively, while the last term is due to the coupled bending-shear interaction at the clamps. For a standard long-string resonator, the length is much greater than the thickness, causing the first term to dominate \cite{sadeghi2020frequency}. In commercial high-stress silicon nitride, the ratio of Young's modulus to stress is approximately 250 \cite{fedorov2020fractal}, leading to a calculated $D_{Q}$ of $\sim 1$. Consequently, the torsion mode cannot experience dissipation dilution \cite{fedorov2019generalized}. In contrast, for a nanoribbon, the width substantially exceeds the thickness as well, resulting in Q factors which scale as $(w/h)^2$ as high as $10^8$ \cite{pratt2023nanoscale}. Provided that curvature at the clamping points is nonnegligible, \eqref{DQ,n} can be reduced to

\begin{equation}
    D_{Q,n} \approx 1 +\frac{1}{\frac{2G}{\sigma}\frac{h^2}{w^2}+\frac{n^2\pi^2}{12}\frac{Eh^2}{\sigma L^2} + \frac{h}{L}\sqrt{\frac{E}{3\sigma}}}.
\end{equation}

The formulation bears strong similarity to the dissipation dilution factor expression for standard string resonators \cite{schmid2016fundamentals,sadeghi2019influence}. This correspondence suggests that established techniques for enhancing the quality factor in flexural modes of string resonators - including but not limited to phononic crystal engineering \cite{beccari2022strained}, tapered clamping structures \cite{bereyhi2019clamp}, and hierarchical mechanical designs \cite{bereyhi2022hierarchical,fedorov2020fractal}- can be effectively extended to enhance the torsion mode performance in ribbon resonators.

\bibliography{references}

\end{document}